\documentclass[twocolumn,prl, aps,superscriptaddress,showpacs]{revtex4-1}

\usepackage{mathptmx}
\usepackage{subfigure}
\usepackage{dcolumn}
\usepackage{amsmath,amssymb}
\usepackage{bm}
\usepackage{color}
\usepackage{latexsym}
\usepackage{epstopdf}
\usepackage{color}
\usepackage[english]{babel}
\usepackage{latexsym}

\usepackage{psfrag,graphicx} 
\usepackage{epsf} 
\usepackage{subfigure} 
\usepackage{amsmath} 
\usepackage{amssymb} 
\usepackage{amsfonts}
\usepackage{bm}
\usepackage{natbib}
\usepackage{epstopdf}
\usepackage{appendix}

\definecolor{mygrey}{gray}{0.35}
\definecolor{myblue}{rgb}{0.2,0.2,0.8}
\definecolor{myzard}{cmyk}{0,0,0.05,0}
\definecolor{mywhite}{rgb}{1,1,1}
\definecolor{myred}{rgb}{1,0.,0.3}

\usepackage[colorlinks=true,citecolor=myblue,linkcolor=myred]{hyperref}

\def\be{\begin{equation}}
\def\ee{\end{equation}}
\def\ba{\begin{align}}
\def\enda{\end{align}}
\def\bi{\begin{itemize}}
\def\ei{\end{itemize}}

 \def\ee{\mathord{\rm e}}

 \def\ee{\mathord{\rm e}}

\newcommand{\Yb}{$^{171}{\rm{Yb}}^{+} $}

\renewcommand{\ee}{{\rm e}}

\def\beq{\begin{equation}}
\def\beq{\begin{equation}}
\def\eeq{\end{equation}}

 \newcommand{\ket}[1]{|#1\rangle}
 \newcommand{\bra}[1]{\langle #1|}

\begin{document}

\title[Short Title]{Simulating the Haldane Phase in Trapped Ion Spins Using Optical Fields}%Quantum Simulation of the Haldane Phase Using Trapped Ions in the Laser-Induced Implementations

\author{I. Cohen}
\affiliation{Racah Institute of Physics, The Hebrew University of Jerusalem, Jerusalem 91904, Givat Ram, Israel}
\author{P. Richerme}
\affiliation{Joint Quantum Institute, University of Maryland Department of Physics and National Institute of Standards and Technology, College Park, MD 20742}
\author{Z.-X. Gong}
\affiliation{Joint Quantum Institute, University of Maryland Department of Physics and National Institute of Standards and Technology, College Park, MD 20742}
\affiliation{Joint Center for Quantum Information and Computer Science, NIST/University of Maryland, College Park, Maryland 20742, USA}
\author{C. Monroe}
\affiliation{Joint Quantum Institute, University of Maryland Department of Physics and National Institute of Standards and Technology, College Park, MD 20742}
\author{A. Retzker}
\affiliation{Racah Institute of Physics, The Hebrew University of Jerusalem, Jerusalem 91904, Givat Ram, Israel}


\begin{abstract}
%{We introduce a quantum simulation scheme of integer-spin antiferromagnetic (AFM) Heisenberg chains ($d=1$) namely, 
%Haldane phase, and higher dimensional frustrated formations ($d>1$) namely, spin liquid phase, in F=1 hyperfine 
%splitting trapped ion platform. We demonstrate the robustness of the ground states to noise in the magnetic field and Rabi 
%frequencies, and propose to detect it by its characterisations: an excitation gap and exponentially decaying 
%correlations, a nonvanishing nonlocal string order, double degenerate entanglement spectrum and the entanglement entropy 
%which obeys the boundary law.}

{We propose to experimentally explore the Haldane phase in spin-one XXZ antiferromagnetic chains using trapped ions. We show how to adiabatically prepare the ground states of the Haldane phase, demonstrate their robustness against sources of experimental noise, and propose ways to detect the Haldane ground states based on their excitation gap and exponentially decaying correlations, nonvanishing nonlocal string order, and doubly-degenerate entanglement spectrum.% By scaling up to higher dimensions and more frustrated lattices, a richer phase diagram is obtained, and the spin liquid phase can be reached.
}
\end{abstract}

\maketitle
\section{introduction}
In a quantum simulation experiment, the behavior of a complex quantum model is examined using controllable system, which acts as the simulator \cite{Feynman,Iulia,Blatt_Roos}. Collections of trapped atomic ions have emerged as excellent standards for quantum simulations of interacting spin models. These electrically charged particles are confined in an electromagnetic trap, micrometers apart from one another \cite{Ion Trap 1,Ion Trap 2}. Thanks to precise spin manipulation capabilities, near-perfect spin readout, and a variety of cooling \cite{cooling1,cooling2} and dynamical decoupling techniques \cite{DD0,DD1,DD2,DD3,DD4,DD5,CDD,Nati,Wang1}, trapped ions can be made to follow target Hamiltonians with high fidelity \cite{fidelity 1,fidelity 2}, making them one of the most promising candidates for quantum simulation. 

To date, trapped ions have mostly been used to simulate spin one-half Hamiltonians, showing the phase transition from the (anti)ferromagnetic to paramagnetic phases in the Ising model \cite{Ising1,Ising2,Ising3,Ising4,Ising5,Ising6,Ising7,Ising8,Edwards,Lin,Islam,Senko} and long range correlation functions in the XX model \cite{xx1,xx2}. By enlarging the spin's degree and moving into integer spin chains, new and subtle physics can appear \cite{Guan1,Guan2,Haldane_conjecture,string,Z2_Z2_1,Z2_Z2_2,symmetry protected topological phase,Itsik,Senko_Haldane}; for example, the local orders vanish and we are left with hidden orders only \cite{hidden review}. 

In last few decades considerable efforts have been made to investigate the non-local physics which appears in integer spin chains, e.g., a spin-one {\it XXZ} antiferromagnetic (AFM) Hamiltonian 
\begin{equation}
H=\sum_{i,j}J_{i,j} \left( S_{x}^{i}S_{x}^{j}+S_{y}^{i}S_{y}^{j}+\lambda S_{z}^{i}S_{z}^{j}\right)+D\sum_{i}\left(S_{z}^{i}\right)^{2}.
\label{HaldaneHamiltonian}
  \end{equation}
 
In 1983, Haldane conjectured \cite{Haldane_conjecture} that Heisenberg chains of integer spins with nearest neighbors antiferromagnetic interactions are gapped, unlike the gap-less half-integer spin chains. This energy gap corresponds to short-range exponentially decaying correlation functions, compared to the long-range power law decaying correlations of half-integer spin systems. Later, Den Nijs and Rommelse \cite{string} suggested that the Haldane phase of the spin one chain is governed by a hidden order, which can be characterized by a nonlocal string order parameter. It is consistent with a full breaking of a hidden $Z_2\times Z_2$ symmetry, which was revealed using a nonlocal unitary transformation by Kennedy and Tasaki \cite{Z2_Z2_1,Z2_Z2_2}. In 2010 Pollmann {\it et al.} \cite{symmetry protected topological phase} showed that the Haldane phase can also be described by the doubly degenerate entanglement spectrum. These characteristics hint that the Haldane phase is a topologically protected phase in one dimension. Hence, quantum simulation of the Haldane phase enables the exploration of topological behavior in relatively simple systems. 

\begin{figure}
\centering
 \includegraphics[width=0.3\textwidth]{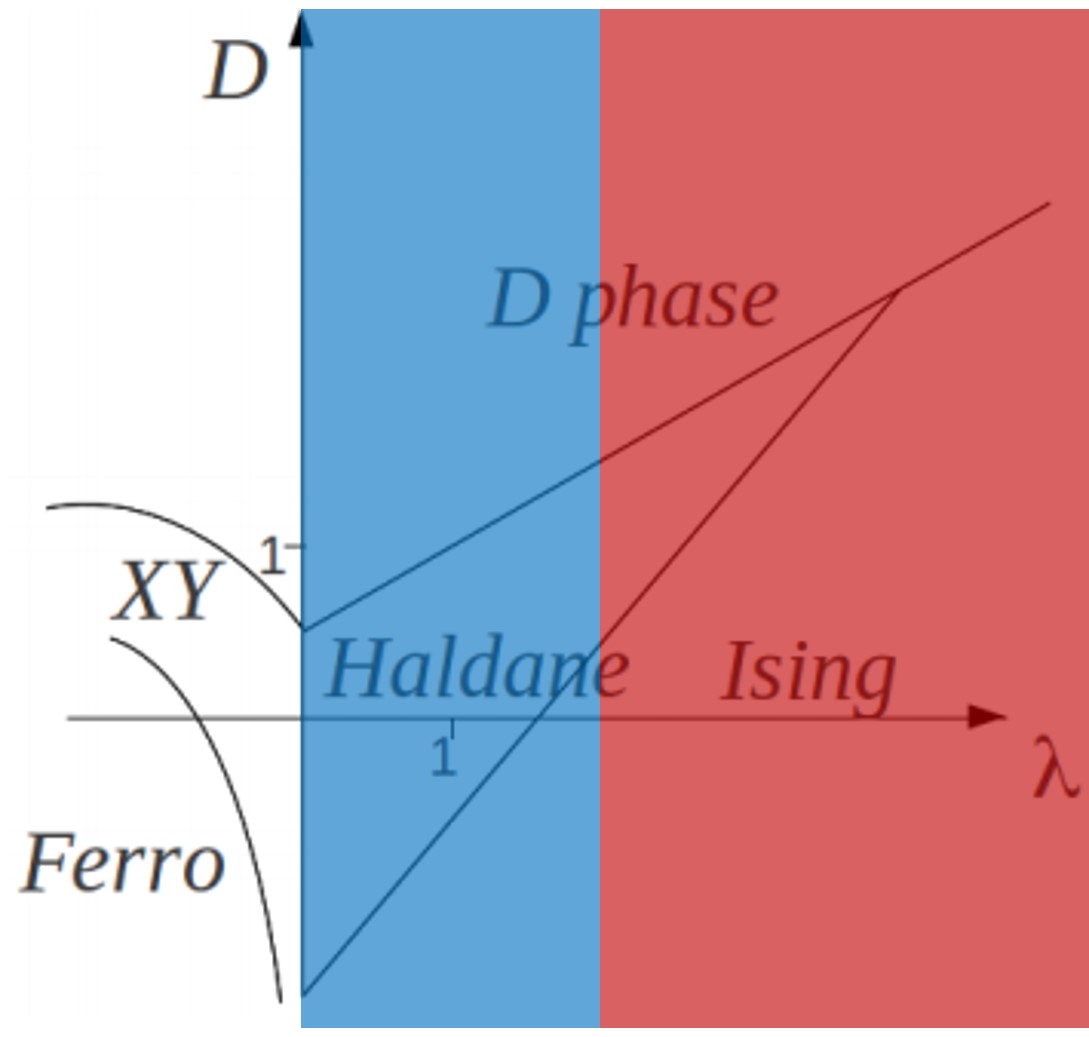}%
  \caption{{\bf The phase diagram of the  spin-one {\it XXZ} antiferromagnetic (AFM) Hamiltonian (eq. \ref{HaldaneHamiltonian}).} The first derivation in this text (as well as the previous derivation \cite{Itsik}) cover $0 \leq \lambda \leq 2$ (in blue), whereas the second derivation in this text covers the whole positive $\lambda \geq 0$ plane (in blue and red).}
\label{phasediagram1}
\end{figure}

Recently an experiment by Senko {\it et al.} \cite{Senko_Haldane} has used trapped ions to simulate spins in higher degrees.
In this experiment a spin-one Hamiltonian was engineered using state-dependent laser forces, paving the way toward highly controllable quantum simulation experiments that exhibit hidden orders. Previously,  we proposed how to engineer the Hamiltonian of Eq. \ref{HaldaneHamiltonian} using microwave-based forces on the trapped ion spins \cite{Itsik}. In this previous derivation of the Hamiltonian, we could cover only $0\leq\lambda \leq 2$ of the phase diagram (Fig. \ref{phasediagram1}). A natural way to extrapolate the microwave approach to a laser based implementation is to use non co-propagating Raman beams instead of a magnetic field gradient \cite{gradient} to induce the spin-phonon coupling.

In this paper, we first propose a new approach to achieving the same result as in \cite{Itsik}, but with a laser based implementation. Then, we propose a new approach for quantum engineering eq. \ref{HaldaneHamiltonian}, while covering the whole positive $\lambda \geq 0$ plane of the phase diagram  (Fig. \ref{phasediagram1}). Instead of having the XX (flip-flop) Hamiltonian $S_y^iS_y^j + S_x^iS_x^j$  as our starting point, in the new approach we engineer a M\o lmer-S\o rensen (MS) like gate \cite{SM} for spin-one systems that takes the form of an Ising Hamiltonian $S_x^iS_x^j$.

We have found three ways to implement the MS gate for spin-one systems \cite{SM,Bermudez gate,Itsik gate}. These schemes were originally proposed for two-level system (qubits), but can be extrapolated to three-level systems (qutrits). For simplicity, we consider the first and second approaches for laser based designs, although a similar derivation exists for the third one, and also for microwave based  implementations.

We explain how to move experimentally to the interaction picture: namely, how to measure in the appropriate basis. Furthermore, we give a detailed explanation of the adiabatic path for reaching the Haldane phase. We stress that our model is robust to the main noise sources in experiments, exploiting dressing fields as a dynamical decoupling technique, and operating in a decoherence free subspace, enabling longer adiabatic evolution times.

\section{Model}
{\it The phonons---} In our model we have $N$ trapped ions each of mass $m$ and electric charge $e$, forming a linear chain along the $z$ axis, $r^0_{i,z}$. The ion formation is determined according to the equilibrium of the Coulomb repulsion between the ions, and the trapping forces. The vibrations of the ions around their settled points, $\Delta r_{i\alpha}$, are solved in the harmonic approximation obtaining the normal modes $ M_{i,n}^{\alpha} $ and $\nu_{n}^{\alpha}$, which are the eigenstates and the eigenvalues of the $n^{th}$ mode and the $i^{th}$ ion in the $\alpha$ direction respectively. Thus, the ion displacements are represented as $\Delta r_{i\alpha}=\sum_{n}M_{i,n}^{\alpha}{\sqrt{\hbar/2m\nu_{n}^{\alpha}}} {\left (b_{n\alpha}^{\dagger}+b_{n\alpha} \right)} $, and the vibration Hamiltonian is represented as $ H_{vib}=\sum_{n,\alpha}\nu_{n}^{\alpha}b_{n\alpha}^{\dagger}b_{n\alpha}$, where $b_{n\alpha}^{\dagger}, b_{n\alpha}$ are the creation and annihilation of a phonon respectively \cite{vib1,vib2,vib3}.

\begin{figure}
\centering
 \includegraphics[width=0.25\textwidth]{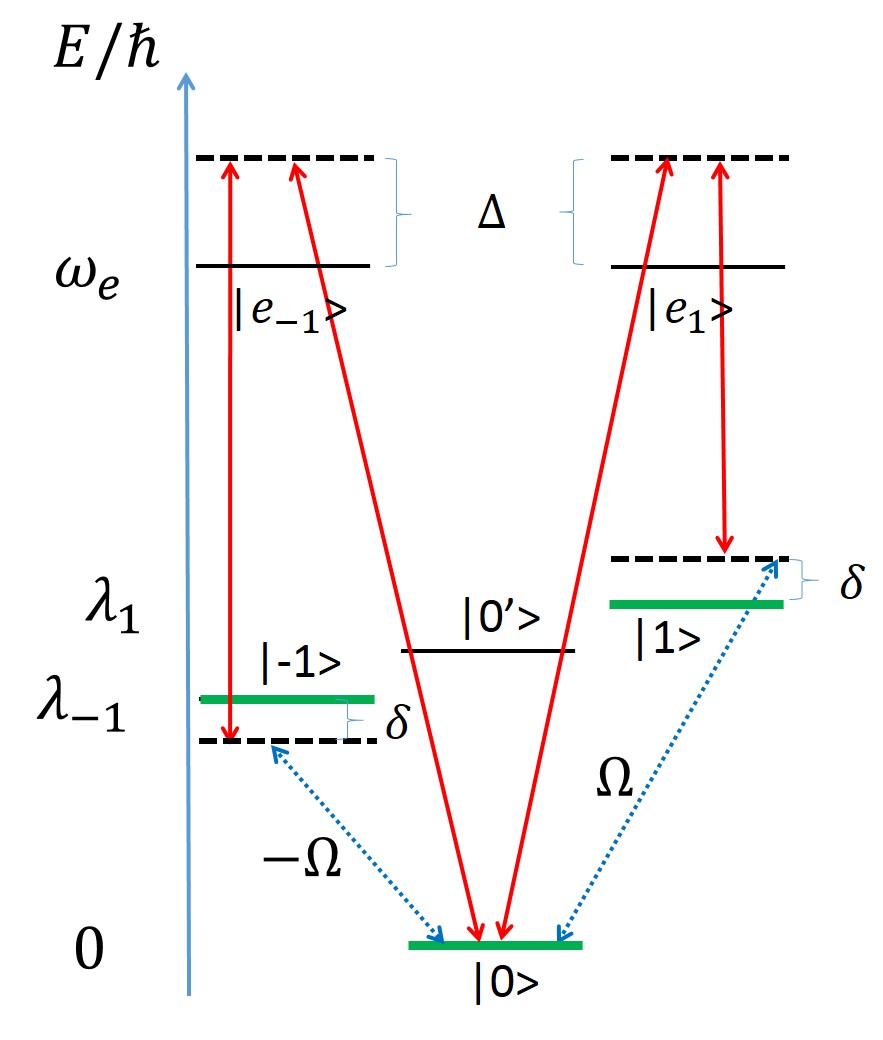}%
  \caption{{\bf  The \Yb ground states used for modeling the spin-one particle (in green).} The beat frequencies between non-co-propagating Raman beams (in red) drive
the $\delta$ detuned transitions between  $\left\vert -1\right\rangle \rightarrow \left\vert 0 \right\rangle $ and $ \left\vert 1 \right\rangle \rightarrow \left\vert 0 \right\rangle $  (in blue), 
generating the spin-laser interaction in eq. \ref{red}).
}
\label{Yb_red}
\end{figure}

{\it The spin---} The second quantum degree of freedom in this model is the spin. For our derivation, we consider the hyperfine structure of the \Yb ion (fig. \ref{Yb_red}), with microwave energy separation between the singlet and the three triplet states \cite{microwave1,microwave2}. Removing the $m_F=0$ state $(\ket{0'})$ from the triplet, we are left with three energy levels for modeling the simulated spin-one system. Quantum manipulation can occur using Raman transitions via virtual excited states, or using microwave driving fields. Note that any other ion having three different energy states in the microwave regime would suffice for modeling the spin-one particle, as long as there is a virtual excited state, through which the Raman transitions between the three levels can occur \cite{monroe-laser}. 

 {\it The interaction---} In trapped ion systems, the spin-spin interaction takes place by exchanging a virtual phonon between the different ions; therefore, a spin-phonon coupling term is needed. Using laser Raman beams, the spin-phonon coupling term is achieved by a sufficiently large Lamb-Dicke parameter of the non co-propagating beams \cite{monroe-laser}.  For antiferromagnetic interactions, we typically choose a radial vibration mode as the mediator between the spins \cite{SS diego,SS monroe}. Therefore, three counter-propagating Raman beams having a momentum difference along the radial direction perform the two beat frequencies corresponding to the $\delta$ detuned transitions between  $\left\vert -1\right\rangle \rightarrow \left\vert 0 \right\rangle $ and $ \left\vert 1 \right\rangle \rightarrow \left\vert 0 \right\rangle $, with a $\pi$ phase difference (fig. \ref{Yb_red}). In the interaction picture with respect to the radial vibration Hamiltonian and the bare state energy structure, we obtain the red sideband transition for a three-level system: 
 \begin{equation}
 \label{red} 
 H_{red}=  \sum_{n,j}\frac{i \Omega\eta_{j,n}}{2\sqrt2} \left( F_+^j  e^{i\delta t} -h.c \right) \left( b_{n}^\dagger e^{i\nu_n t} +h.c \right).
\end{equation}
After applying the Lamb-Dicke approximation $\eta_{j,n}\sqrt{N_n+1}\ll1$, where $N_n$ is the average number of phonons of the $n^{th}$ mode, $\eta_{j,n}=k_L M_{j,n}\sqrt{\hbar/2m\nu_n }$ is the Lamb-Dicke parameter, $k_L$ is the laser's wavenumber, and $F_+^j=\sqrt2\left( \left\vert1\right\rangle^j \left\langle 0 \right\vert + \left\vert 0\right\rangle^j \left\langle -1 \right\vert  \right).$ 
In the second-order perturbation approach, this results in a XX Hamiltonian %$ F^j_+ F^i_- + h.c $ 
in addition to a residual term \cite{Senko_Haldane}:
     \begin{equation}
\begin{split}
     \label{XXHamiltonian}
 H_{XX}=\sum_{i\le j}J_{ij}^{eff}\left(\left(F_{x}^{i}F_{x}^{j}+F_{y}^{i}F_{y}^{j}\right)\left(1-\delta_{i,j}\right)-
 \frac{\left(F_{z}^{j}\right)^{2}}{2}\delta_{i,j}\right),\\
 H_{res}=\sum_{j,n,m}J_{jnm}^{res}F_{z}^{j}\left(b_{n}^{\dagger}b_{m}+
 \frac{1}{2}\delta_{n,m}\right)
 e^{-i\left(\nu_{n}-\nu_{m}\right)t},
\end{split}
   \end{equation}
 where
\begin{equation}
\begin{split}
J_{ij}^{eff}=\sum_{n} \frac{\eta_{i,n}\eta_{j,n}}{2}  
\frac{ \nu_{n}\Omega^2}{\delta^{2}-\nu_{n}^{2}}
\propto\left|
\vec{r_{i}^{0}}-\vec{r_{j}^{0}}\right|^{-\xi}_{i\neq j},\\
J_{jnm}^{res}=\Omega^2 \delta\frac{\eta_{j,n}\eta_{j,m}}{4}  
\left(\frac{1}{\delta^{2}-\nu_{n}^{2}}+
\frac{1}{\delta^{2}-\nu_{m}^{2}}\right).
\end{split}
 \label{XXcoefficients}
\end{equation}
Our model can be decoupled from the residual term if $J_{jnm}^{res}$ is approximately uniform across the chain, such that $H_{res}\propto \sum_i F_z^i$. This is due to the fact that the model's relevant ground states, as we will see later in this paper, have a vanishing projection over the z axis, thus the experiment is performed in the decoherence-free subspace. As was suggested in Ref. \cite{Senko_Haldane}, this residual term can be eliminated by adding additional beat frequencies that generate the blue sideband transitions together with the red sideband ones, such that the MS transitions for spin-one systems is realized.

Moreover, using trapped ions, the spin-spin interaction is not the nearest-neighbor-only as the classic Haldane work has considered, %in eq. \ref{HaldaneHamiltonian}, 
but rather a power-law decay \cite{SS diego,SS monroe}. However, as Ref. \cite{Gorshkov,Gorshkov2} shows, the Haldane phase can still be found in power-law interacting systems.

%In our previous derivation \cite{Itsik prl}, we have used eq. \ref{XXHamiltonian} as our starting point, resulting in a limited covering of the phase diagram. We will show here that much better can be done.  
 Up to this stage of the derivation of the AFM {\it XXZ} Hamiltonian, we have shown how to generate the first two terms of eq. \ref{HaldaneHamiltonian}. We now pursue the two other tunable terms, the $D$ and $\lambda$ terms.

\begin{figure} 
\centering 
 \includegraphics[width=0.5\textwidth]{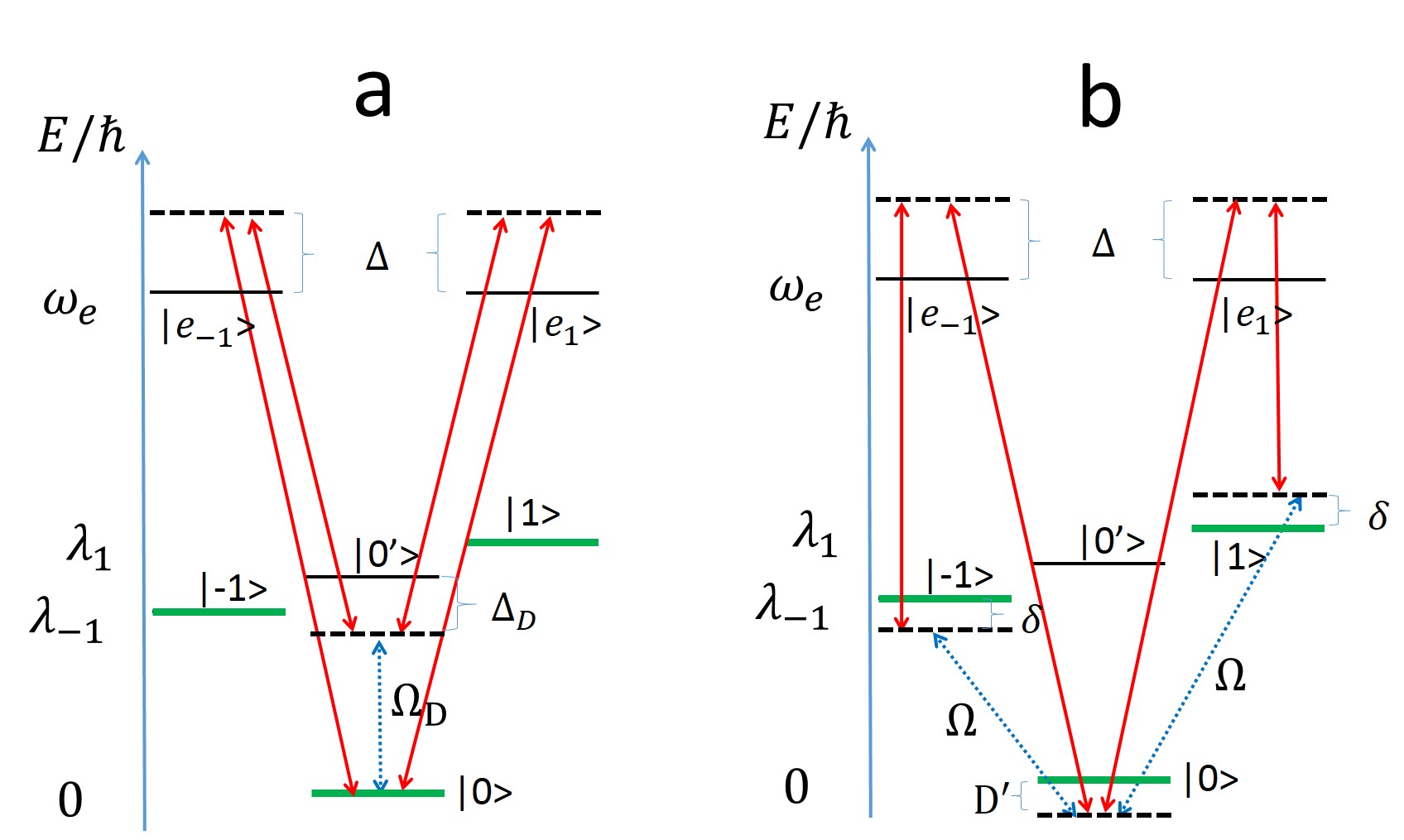}
  \caption{{\bf Generating the $D$-term} Generating the $D$-term in eq. \ref{HaldaneHamiltonian} is done in two ways: (1) (a.) Applying two additional co-propagating Raman beams, corresponding to a $\Delta_D$ detuned transition between $\left|0'\right\rangle\longleftrightarrow\left|0\right\rangle$ results in an A.C. Stark shift $\frac{\Omega_D^2}{4\Delta_D}\left\vert 0\right\rangle \left\langle 0 \right\vert $. This could alternatively be generated using a $\Delta_D$ detuned microwave driving field. (2) (b.) Within the non-copropagating Raman beams (red) (fig. \ref{Yb_red}), we regard the $\left\vert 0 \right\rangle$ state as $D'$-shifted, and move to the interaction picture with respect to the $D'$-shifted bare energy structure. Thus, we are left with $-D'\left\vert 0 \right\rangle \left\langle 0 \right\vert$, which is $D'F_z^2 $. In our later derivation (which covers the entire $\lambda \geq 0$ plane), we will impose a $2D'$ detuning rather than a $D'$ detuning.}
 \label{D_fig_1}
\end{figure}

{\it Generation of the control anisotropy D-term ---} Generating the tunable D-term can be done in two ways: (1) using an additional transition to generate the A.C. Stark shift, from which the D-term will arise (Fig. \ref{D_fig_1} a); (2) imposing detunings on all the previous driving fields,such that the D-term is set aside from the bare state energy structure (Fig. \ref{D_fig_1} b). To generate the A.C. Stark shift we can apply an additional $\Delta_D$ detuned microwave driving field, corresponding to the transition between the two zero states of the triplet and the singlet, $\left|0'\right\rangle\longleftrightarrow\left|0\right\rangle$ (fig.\ref{D_fig_1} a). Similarly we can obtain the same result by applying co-propagating Raman beams corresponding to the same transition. Thus we obtain:
\begin{equation}
\frac{\Omega_D}{2} \left\vert 0 \right\rangle \left\langle 0' \right\vert e^{i \Delta_D t }+ h.c,
\label{A.C}
\end{equation} 
which yields an A.C. Stark shift term $D'(F_z)^2$ with $D'=-\frac{\Omega_D^2}{4\Delta_D}$,  in the second order perturbation approach, assuming $ \frac{\Omega_D}{2} \ll \Delta_D $.

Generating a tunable $D$-term can alternatively be achieved by regarding the $\left|0\right\rangle$ bare state energy level as $D'$-shifted, corresponding to its real value in the hyperfine structure (fig.\ref{D_fig_1} b). In other words, we impose $D'$-detuning on all the driving fields generating the XX Hamiltonian (eq. \ref{XXHamiltonian}). Therefore, we are left with the term $-D' \left\vert 0\right\rangle \left\langle 0 \right\vert=D'(F_z)^2$. In order to tune this term we have to shift all the driving fields transitions continuously, as was recently shown in \cite{Senko_Haldane}. Note that the anisotropy $D$-term is slightly different from these $D'$ terms, since the model is $\theta$ rotated while generating the $\lambda$-term, as is explained below. 

\begin{figure} 
\centering 
 \includegraphics[width=0.5\textwidth]{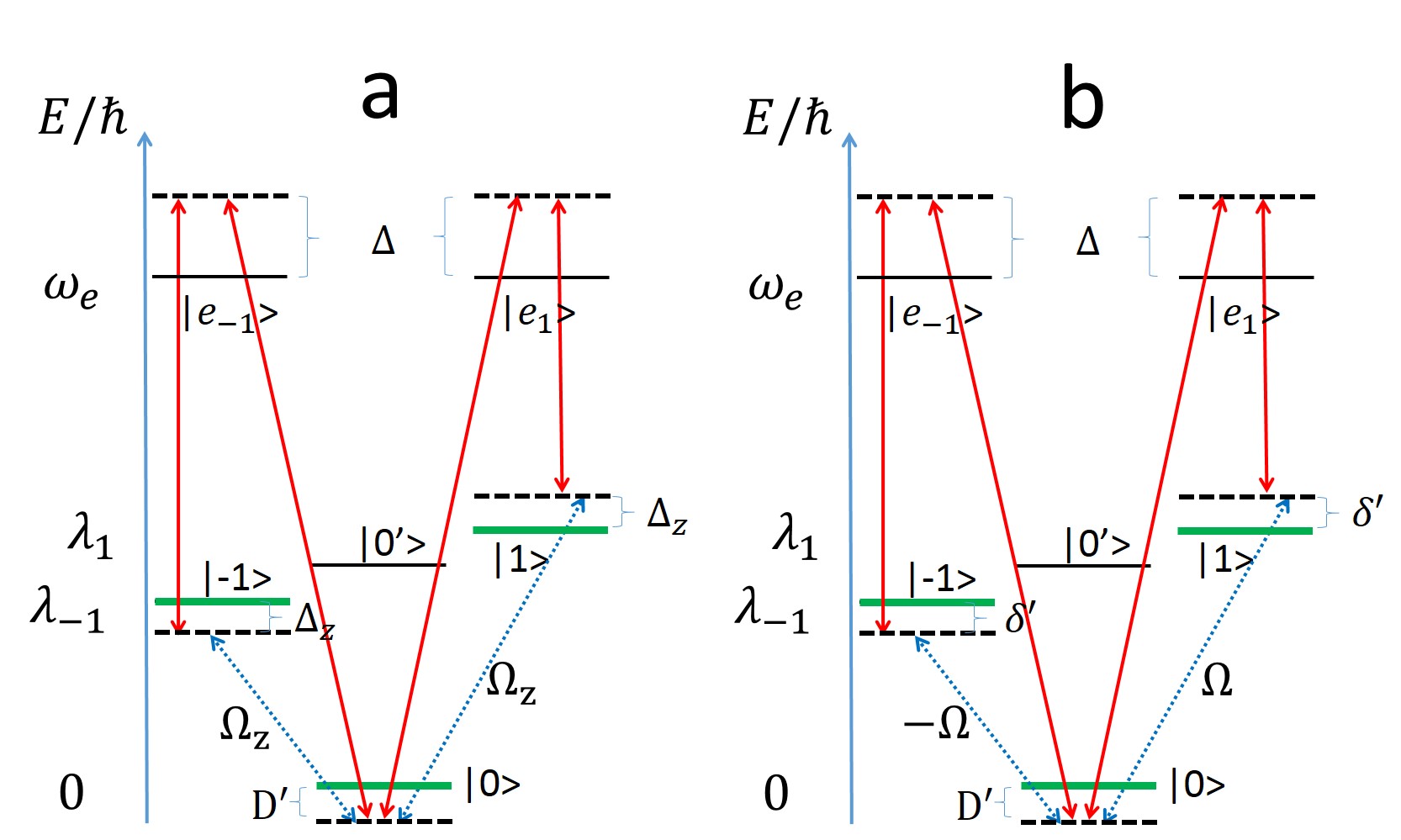}
  \caption{{\bf Generating $\Omega'F_z\cos\theta$ for the $\lambda$-generating trick.} There are two ways of generating this term: (1) (a.) Additional copropagating Raman beams (red) perform the $\pm\Delta_z$ detuned transitions between $\left|\pm1\right\rangle \leftrightarrow\left|0\right\rangle$ respectively, with a Rabi frequency of $\Omega_z$. These effective transitions can alternatively be realized using detuned microwave driving fields (blue). Therefore, we obtain the desired term from the A.C. Stark shifts where $\Omega'\cos\theta=\Omega_{z}^2/\Delta_z$. (2) (b.) Adding detunings to the non-copropagating Raman beams (red) (fig. \ref{Yb_red}), such that $\delta'=\delta+ \Omega'\cos\theta$. In this way we are left with the desired term, after moving to the interaction picture with respect to the detuned bare energy structure.}
 \label{F_z1}
\end{figure}

  {\it Generation of the Ising-like $\lambda$-term ---} The Ising-like anisotropy term $\lambda J_{i,j} F_{z}^{i}F_{z}^{j}$ is produced using a technique, which is revealed by considering the interaction picture. It is done by adding a spin operator term, $\theta$-rotated from the $z$ axis, namely, $\Omega' F_{z,\theta}=\Omega' \left( F_z \cos\theta +F_{\alpha}\sin\theta \right)$, where $\alpha$ can be either $x$ or $y$ or their superposition (XY plane), and generating each component separately. $\Omega' F_{z}\cos\theta$ can be produced similarly to the two ways that the D term was generated. The first way is to drive the transitions between  $\left|\pm1\right\rangle \leftrightarrow\left|0\right\rangle$ with  $\pm\Delta_z$ detunings respectively. These transitions can be generated using co-propagating Raman beams, or alternatively using microwave driving fields, with a Rabi frequency $\Omega_{z}$, such that the effective A.C. Stark shifts result in the $\Omega' F_{z}\cos\theta$ term, with $\Omega'\cos\theta=\Omega_{z}^2/\Delta_z$ (Fig. \ref{F_z1} a). The second way to induce $\Omega' F_{z}\cos\theta$  is by imposing detunings on the non co-propagating Raman beams; i.e., by imposing $\pm\Omega'\cos\theta$ detunings on the transitions between $\left|\pm1\right\rangle \leftrightarrow\left|0\right\rangle$ respectively, such that the $F_{z}\cos\theta$ term is set aside from the bare state energy structure (Fig. \ref{F_z1} b). 

\begin{figure} 
\centering 
 \includegraphics[width=0.5\textwidth]{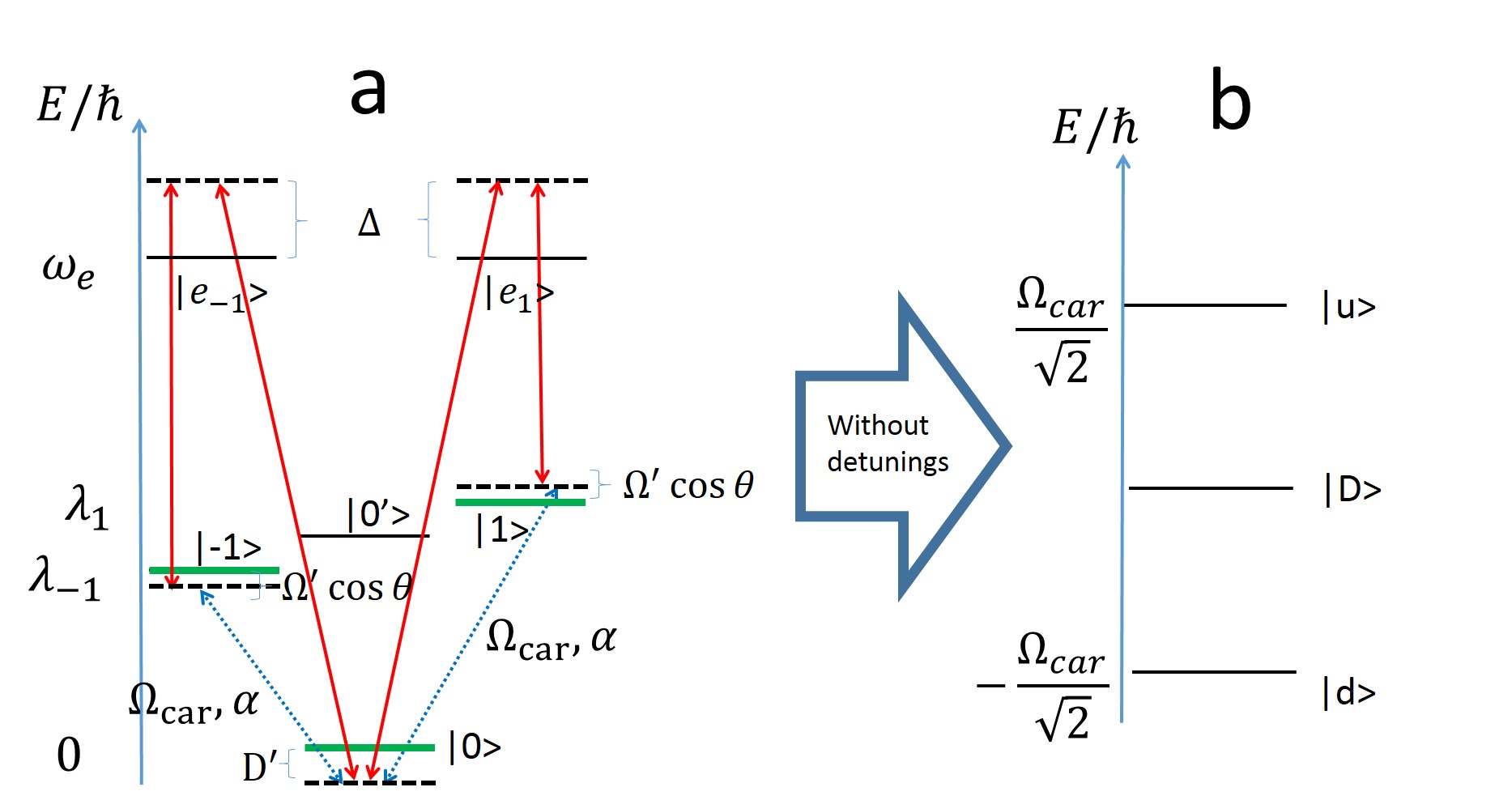}
  \caption{{\bf Generating $\Omega'F_{\alpha}\sin\theta$ for the $\lambda$-generating trick.} (a.) We apply carrier transitions using additional copropagating Raman beams (red), or alternatively microwave driving fields (blue), while keeping the imposed detunings for generating the D-term (fig. \ref{D_fig_1} a) and the $\Omega'F_z\cos\theta$ term (fig. \ref{F_z1} a). Here,  $\Omega_{car}=\Omega'\sin\theta\sqrt2$, and $\alpha=\alpha_1-\alpha_{-1}$ is determined by the initial phase difference of the two effective transitions. The same carrier transitions, only without detunings, are used in our later approach as dressing fields that transform the system into the dressed state basis (b.), thus protecting it from the magnetic noise. }
 \label{F_a1}
\end{figure}

The second term $\Omega'F_{\alpha}\sin\theta$ is obtained using the carrier transitions, where their relative phase difference determines the resulting operator direction $\alpha$. The carrier transitions can be applied either using co propagating laser Raman beams, or simply using microwave driving fields, with a Rabi frequency $\Omega_{car}=\Omega'\sin\theta\sqrt2$  (fig. \ref{F_a1} a).
 
By applying a $\theta$ rotation around the perpendicular axis, such that $F_{z,\theta}$ is transformed into $\overline{F_{z}}$, the new spin operators and the new basis are $\theta$ rotated as well.   %the operators are transformed to: $S_{z,\theta}\rightarrow S_{z}$, $S_{z}\rightarrow S_{z}\cos\theta+S_{x}\sin\theta$, $S_{x}\rightarrow S_{x}\cos\theta-S_{z}\sin\theta$, $S_{y}\rightarrow S_{y}$, and the new dressed-state basis is $\theta$ rotated as well. 
  If we move to the interaction picture with respect to  the new term we have built $\Omega'\overline{F_{z}}$, and use the RWA where we require $\frac{(\Omega\eta_{j,n})^2}{8(\delta-\nu_n)}\ll\Omega' \ll \delta-\nu_n $, we end up with the following effective Hamiltonian: 
  \begin{equation}
  \begin{split}
\label{Heff_final}
 H_{eff}=\sum_{i < j}{J_{ij}^{eff}}\left\lbrace\left( \overline{F_{x}^{i}}\overline{F_{x}^{j}}+\overline{F_{y}^{i}}\overline{F_{y}^{j}} \right)\frac{1+\cos^2\theta}{2} + \overline{F_{z}^{i}}\overline{F_{z}^{j}}\sin^2\theta\right\rbrace\\
+\sum_i\left(D'-\frac{J_{ii}^{eff}}{2}\right)\left(\frac{\cos^2\theta}{2} -\sin^2\theta \right)(\overline{F_z^i})^2.
 \end{split}
 \end{equation}
We have obtained the required spin-one  {\it XXZ} AFM Hamiltonian, where we can only cover $0\leq\lambda \leq 2$. Experimentally, the following parameters are realistic: $\nu_n=2\pi  \cdot 5$MHz, $\delta=2\pi \cdot 5.1$MHz, $\Omega=2\pi  \cdot 500$kHz,$\eta_{n,j}\approx 0.14$, $\Omega'=2\pi  \cdot  10$kHz, which gives $J_{i,i+1}^{eff} \approx 2\pi  \cdot  1$kHz, such that $0<D<2\pi  \cdot  10$kHz.

An important advantage over the previous derivation in \cite{Itsik}, is that in order to obtain the antiferromagnetic interactions, $\delta$ should be larger than the secular frequencies of the radial vibrational modes, rather than the Rabi frequency $\Omega/\sqrt{2}$. Since one of the main experimental challenges is to reach a high Rabi frequency, in the present derivation we overcome this technical obstacle.  As for the residual terms we dropped on the way, such as the carrier transition that yields a four-photon A.C. Stark shift and was neglected from eq. \ref{red}, and the neglected fast rotating terms and the residual from eq. \ref{XXHamiltonian}, these terms operate with $\sum_j\overline{F_{z}^{j}}$. Since the model's relevant ground states have a vanishing projection over the $z$ axis, we are decoupled from these undesired terms.

In this scheme, we move more than once to different interaction pictures. In the following section, we explain how this can be done experimentally.

%\begin{figure} 
 % \centering
 %\includegraphics[width=0.5\textwidth]{4.eps}
 % \caption{{\bf The additional driving fields in the first new derivation }  Right side: Applying a radio-wave driving field, on resonance with the dressed-state energy structure, for two reasons: 1) To suppress the limiting red side-band transition and to obtain the SM $S_z^iS_z^j$ interaction instead. 2) To achieve the second term $\Omega'\sin\theta F_z$ of the $\lambda$-generating trick. Left side: For the first term $\Omega' \cos\theta F_z$ of the $\lambda$-generating trick, we apply additional four co-propagating Raman beams (red color), performing the two $\delta$-detuned Raman transitions between $\left\vert -1\right\rangle \rightarrow \left\vert 0 \right\rangle $ and $ \left\vert 1 \right\rangle \rightarrow \left\vert 0 \right\rangle $ (blue color), where the spin-phonon coupling term is suppressed.}
%   \label{ND1}
%\end{figure}

\section{moving to the interaction picture experimentally}
Since the measurement is taken in the lab frame and the effective Hamiltonian is derived after moving to several interaction pictures, we now show how the dynamics of the state in the lab frame relates to the dynamics determined by the last interaction picture. For this purpose, we take the general case where in the lab frame the evolution of the Schrodinger state is described as follows:
\begin{equation}
\begin{split}
H_s(t)=H_0+H_{int}(t)\\
U_s(t)=T\int_0^t \exp \left(-i H_s (t')dt' \right)\\
\left\vert\Psi_s(t)\right\rangle = U_s(t)\left\vert\Psi_s(0)\right\rangle
\end{split}
\end{equation} 
where $T$ is the time ordering operator. Moving to the interaction picture, the dynamics is described as:
\begin{equation}
\begin{split}
U_0^\dagger(t)=\exp \left(i H_0 t \right)\\
H_{I_0}(t)=U_0^\dagger(t) H_{int}(t) U_0(t)\\
U_{I_0}(t)=T\int_0^t \exp \left(-i H_{I_0} (t')dt' \right)\\
\left\vert\Psi_{I_0}(t)\right\rangle = U_{I_0}(t)\left\vert\Psi_{I_0}(0)\right\rangle,\\
\end{split}
\end{equation}
Usually, the first interaction picture $H_0$ is time-independent since it is the bare state energy structure. However, in general, $H_0$ can also be time-dependent, where $U_0^\dagger(t)=\exp \left(i \int_0^t H_0(t') dt' \right)$. Here, we only assume that $H_0$ is a single particle operator or a global rotation; i.e., it does not create entanglement. In fact, if the realization of the $D$ control parameter is achieved using detunings, varying it during the Haldane phase simulation experiment effectively results in having the first interaction picture $H_0(t)$ time-dependent.

Using the definition of the interaction picture
\begin{equation}
\left\vert\Psi_{I_0}(t)\right\rangle = U_0^\dagger(t)\left\vert\Psi_s(t)\right\rangle,
\end{equation}
we obtain the relation between the interaction and the Schrodinger pictures:  
\begin{equation}
 U_{I_0}(t)=U_0^\dagger(t)U_s(t),
\label{Interaction0}
\end{equation}
since $\left\vert\Psi_{I_0}(0)\right\rangle=\left\vert\Psi_s(0)\right\rangle$.

%In our proposal (spin-$1$ systems), as in the spin-$1/2$ systems (Lieb–Robinson paper), we move to the interaction picture once again. Namely,
Suppose that the first interaction frame is by itself a "Schrodinger" frame for the next interaction picture, such that:
\begin{equation}
H_{I_0}(t)=H_1(t)+H_{I_0}^{int}(t),
\end{equation}
where $H_1(t)$ is a single particle operator in the first interaction frame. As before 
\begin{equation}
\begin{split}
U_1^\dagger(t)=\exp \left(i \int_0^t H_1(t') dt' \right)\\
H_{I_1}(t)=U_1^\dagger(t) H_{I_0}^{int}(t) U_1(t)\\
U_{I_1}(t)=T\int_0^t \exp \left(-i H_{I_1} (t') \right)dt'\\
\left\vert\Psi_{I_1}(t)\right\rangle = U_{I_1}(t)\left\vert\Psi_{I_1}(0)\right\rangle,\\
\end{split}
\end{equation}
thus, 
\begin{equation}
U_{I_1}(t)={U_1}^\dagger(t)U_{I_0}(t).
\label{Interaction1}
\end{equation}
Substituting eq. \ref{Interaction0} in eq. \ref{Interaction1} we obtain
\begin{equation}
U_s(t)=U_0(t)U_1(t)U_{I_1}(t); \quad \left\vert\Psi_s(t)\right\rangle=U_0(t)U_1(t) \left\vert\Psi_{I_1}(t)\right\rangle
\end{equation}
For $N+1$ interaction pictures in our derivation, we obtain
\begin{equation}
\begin{split}
U_s(t)=U_0(t)U_1(t)...U_N(t)U_{I_N}(t)\\
\left\vert\Psi_s(t)\right\rangle=U_0(t)U_1(t)...U_N(t) \left\vert\Psi_{I_N}(t)\right\rangle.
\end{split}
\end{equation}
In order to move experimentally to the interaction picture, such that the Schrodinger state will evolve according to the last interaction Hamiltonian, we have to rotate the system back to counter $U_0(t)U_1(t)...U_N(t)$. 
There are two main ways to move experimentally to the interaction picture, as will be discussed next.

The first and straightforward way is by applying $N$ concatenated global rotations for each interaction picture, except for the first one%, assuming that every interaction unitary $U_k(t)$ is generated by different driving fields, which can be blocked from the system
. Suppose we want to measure the system after time $\tau$, in which all the driving fields were on. In order to counter the last interaction unitary $U_N(\tau)$, we only block the driving fields that generate the simulated Hamiltonian $U_{I_N}(\tau)$, while operating with all the driving fields that are responsible for the interaction pictures. This will be done for time $t_N$, such that we obtain:
\begin{equation}
U_s(\tau+t_N)=U_0(\tau+t_N)U_1(\tau+t_N)...U_N(\tau+t_N)U_{I_N}(\tau),
\end{equation} 
and $t_N$ is determined by $U_N(\tau+t_N)=\mathbb{I}$. The same approach can be used to counter all the other interaction unitary,
%Countering $U_k(t)$ after all $U_n(t)$ for $k<n\leq N$ have already been countered during $\sum_{n=jk+1}^N t_n$, (with the same approach taken for countering $U_{I_N}(t)$), is done in the following way. we can block all the driving fields that are responsible to higher order interaction pictures in addition to the simulated Hamiltonian, namely $U_{Ik}(t)$. This will be done for time $t_k$, thus, $U_s(t+t_k)=U_1(t+t_k)...U_k(t+t_k)U_{Ik}(t)$, determining $U_k(t+t_k)=1$. 
%Concatenating this global rotation $N$ times, for the $n$ interaction unitary can be achieved using the following pulse sequence:
obtaining the following pulse sequence:
\\(1). All the driving fields for the experiment time $\tau$.
\\(2). All the driving fields that are responsible for the $H_1(t)$ to $H_N(t)$ interaction pictures for $t_N$ time.
\\(3). All the driving fields that are responsible for the $H_1(t)$ to $H_{N-1}(t)$ interaction pictures for $t_{N-1}$ time.
\\\vdots
\\(N). Only the driving fields that are responsible for the $H_1(t)$ interaction pictures for $t_1$ time. 

In that way, we obtain the following relation between the last interaction picture and the Schrodinger one:
\begin{equation}
\begin{split}
U_s(\tau+t_N+...+t_1)=\\
U_0(\tau+t_N+...+t_1)U_1(\tau+t_N+...+t_1)...U_N(\tau+t_N)U_{I_N}(\tau).
\end{split} 
\end{equation}
Setting $U_k(\tau+t_N+...+t_k)=\mathbb{I}$ for all $1\leq k\leq N$, the Schrodinger state therefore evolves according to the last interaction picture
\begin{equation}
U_s(\tau+t_N+...+t_1)=U_0(\tau+t_N+...+t_1)U_{I_N}(\tau).
\end{equation}
with an additional phase $U_0(\tau+t_N+...+t_1)$, which we do not measure. 
If we want to measure in any different basis, we can rotate the system using any one of $U_k(\tau+t_N+...+t_k)=e^{i\sigma_{\alpha_k}\theta_k}$.

%The second way to move experimentally to the interaction picture is once again using pulse sequence, to counter each $U_k(t)$ global rotation in the reversed order to the first way, namely, starting to counter the second interaction unitary $U_1(t)$ until $U_n(t)$.  Therefore, the pulse sequence is as follows:
%\\1. All the driving fields for the experiment time $\tau$, obtaining $U_s(\tau)=U_0(\tau)U_1(\tau)...U_n(\tau)U_{In}(\tau)$.
%\\2. Only the driving fields that are responsible for the $H_1(t)$ interaction pictures for $t_1$ time, obtaining $U_s(\tau+t_1)=U_0(\tau+t_1)U_1(\tau+t_1)...U_n(\tau)U_{In}(\tau)$, such that $U_1(\tau+t_1)=1$.
%\\3. All the driving fields that are responsible for the $H_1$ to $$H_{n-1}$ interaction pictures for $t_{n-1}$ time.
%\\...
%\\n. Only the driving fields that are responsible for the $H_1$ interaction pictures for $t_1$ time. 

The second way to move experimentally to the interaction picture is simpler. Since the pre-factor of $U_{I_N}(\tau)$ is a global rotation $U_0(\tau)U_1(\tau)...U_N(\tau)=e^{i\sigma_{\alpha_{tot}}\theta_{tot}}$, we can apply just one global rotation to counter the total interaction unitary rotations all together. Since any rotation in the Bloch sphere can be generated by two independent rotations (e.g. two orthogonal rotations), it is sufficient to use the first and second interaction pictures, namely $H_0(t)$ and $H_1(t)$, which are orthogonal rotations for that task. Measuring in any other basis can be achieved by an additional rotation, which can be added to all the interaction unitary, and thus be represented effectively with $H_0(t)$ and $H_1(t)$. Next, we explicitly show how to apply the first approach in the two derivations of the {\it XXZ-D} Hamiltonian.

%It means that after the experiment was done, we have to operate two global rotations: the first $U^1_{0}(\tau)$ in the interaction picture, and the second $U_0(\tau)$ is in the lab frame. In order to generate a global rotation in the interaction picture, we have to apply a driving field that will be time-independent in the interaction frame. If the second interaction picture is a a "Schrodinger" frame for the next interaction picture, we just have to operate another global rotation, only in the next interaction frame, and so on. 

%It is easy to say theoretically that we can apply two global rotations to counter each of $U_0(t)U_1(t)$ separately. Namely we can shut all our driving fields down, and let the system evolve freely only with $H_0$ for a certain time $\tau_0$ that compensates the phase accumulated due to $U_0(t)$. Then, we can apply a rotation 
% however, there is an experimentally easier way to yield the same result. For that we have to apply a  time independent rotation in the last interaction frame, where the driving field have already accumulated the correct time dependent phases. 
 
{\it Experimental realization---}
While generating the {\it XXZ-D} Hamiltonian we move twice to different interaction pictures: the first one is with respect to the bare energy gap of the qutrit, yielding the following interaction unitary:
\begin{equation}
U_0(\tau)=\exp\left(-i\left[(\omega_0-D') \left\vert 0 \right\rangle \left\langle 0 \right\vert  + \lambda_1\left\vert 1 \right\rangle \left\langle 1 \right\vert -\lambda_1\left\vert -1 \right\rangle \left\langle -1 \right\vert  \right]\tau \right)  
\label{U1}
\end{equation} 
where we used the detuning approach to generate the $D$ term. The second interaction picture is with respect to the $\lambda-$generating term, resulting in the following interaction unitary:
\begin{equation}
U_1(\tau)=\exp\left(-i \Omega'\left[\cos\theta F_z+\sin\theta F_\alpha \right] \tau\right),
\end{equation}
with $\alpha=x,y$.
Moving to these interaction pictures results in the effective Hamiltonian (eq. \ref{Heff_final}) which yields the simulated evolution $U_{I_1}(\tau)$.

As was discussed above, in order to move to the interaction frame, we first have to shut down the effective Hamiltonian. This is done by blocking the two non-copropagating Raman beams (setting $\Omega=0$), and eliminating the $D'$ term. %Recall that we have suggested two ways to generate this, either by using a $\Delta_D-$detuned transition that results in an A.C. Stark shift resulting in the $D'$ term (eq. \ref{A.C}), or by imposing a $D'-$detuning to all the relevant driving fields. Therefore, we can block the driving fields that generate this term with an A.C. Stark shift, or cancel the $\pm\Omega'\cos\theta$ detunings of the carrier transitions, if the second approach is taken. 
Shutting down the effective Hamiltonian should take $t_1$ such that $U_1(\tau+t_1)=\mathbb{I}$.
 
There is no need to counter $U_0$ since it only yields an unmeasured phase. Leaving the experiment at this point results in measuring the state in the $F_z$ basis. Now, if we want to measure in the simulated basis, namely $\overline{F_{z}}= \cos\theta F_z+\sin\theta F_\alpha $ for $\alpha=x,y$, we need to rotate the system around a perpendicular axis. Since by using the carrier transitions we generate $\Omega'\sin\theta F_\alpha$, we have to rotate the system with an $F_\beta$ operation, with $\beta \perp \alpha$. This can be achieved after countering $U_1$, by a $\pi/2$ phase change of the laser or microwave driving field, such that instead of operating with $F_x$ the driving field will operate with $F_y$. In a similar way we can measure in any basis we choose.

\section{adiabatic quantum simulation} 
Once we know how to quantum engineer the Hamiltonian of the system under investigation with tunable parameters, the adiabatic quantum simulation proceeds as explained below. We initialize the system in a trivial phase by appropriately setting the parameters of the Hamiltonian, and we initialize the system's state in its trivial ground state. Then, we change the Hamiltonian parameters adiabatically, slower than the energy gap, such that the system stays in the ground state of the instantaneous Hamiltonian, until it reaches the ground state of the non-trivial phase that we want to investigate.

In our case the system under investigation is the Antiferromagnetic XXZ-D spin-1 Hamiltonian, and its trivial phase is the large-D phase, with a tensor product of $\left\vert 0 \right\rangle$ in each site, as its trivial ground state. Since we are operating with a rotated basis $\overline{F_{z}}=\cos\theta F_z+\sin\theta F_\alpha $, we can initialize the system in a tensor product of $\left\vert 0 \right\rangle$ in the $F_z$ basis using polarization  \cite{microwave1,microwave2}. Then we can rotate the state with orthogonal operations: $F_\beta$ (with $\alpha \perp \beta$), similarly to what was described above. Now, we set a large $D$ parameter of the Hamiltonian, and start the adiabatic variation of the $D$ parameter until we reach the Haldane phase regime.  
  
\begin{figure} 
\centering 
 \includegraphics[width=0.25\textwidth]{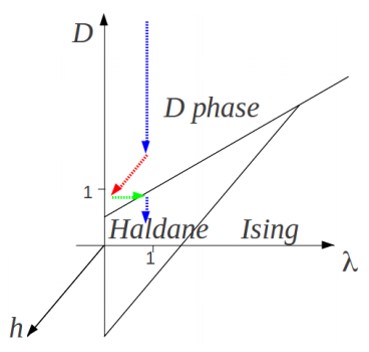}
  \caption{{\bf Adiabatic path using the symmetry breaking perturbation.} When crossing a second order phase transition the energy gap is closed, and the adiabatic approximation is invalid. We break the symmetries of the Hamiltonian (eq. \ref{HaldaneHamiltonian}) and thus, go around the phase transition, keeping a finite energy gap during the whole path.}
 \label{adiabatic}
\end{figure}

When getting closer to the thermodynamic limit, the energy gap in the second order phase transition closes \cite{Sachdev}; thus the adiabatic approximation cannot hold for increasingly long chains. Overcoming this problem takes advantage of the fact that the Haldane phase is a symmetry-protected topological phase. Thus, we can use a symmetry-breaking perturbation in order to go around the $D\rightarrow H$ phase transition, while still operating adiabatically (Fig. \ref{adiabatic}).
  
The Haldane phase is protected by the following symmetries: a bond centered spatial inversion $\vec{S}_{j}\rightarrow\vec {S}_{-j+1}$, a time reversal symmetry $\vec{S}_{j}\rightarrow-\vec{S}_{j}$ and the dihedral $D_{2}$ symmetry, which is a $\pi$ rotation around the $x$, $y$ and $z$ axes. In order to break all these symmetries we add a perturbation term $H_{pert}=-h\sum_{i}\left(-1\right)^{i}S_{z}^{i}$. Since we know how to engineer the $S_z$ term, which is $\overline{F_{z}}$ in our derivation, the symmetry breaking term can be produced by individual addressing, which is achieved by focusing the laser beams.
 
Note that since the quantum simulation experiment is adiabatic, its duration has a lower bound determined by the energy gap. However, like any other quantum experiment, it also has an upper bound that is determined by the coherence time. If the experiment lasts longer than the coherence time, decoherence processes might destroy the quantum information carried by the system, and the results cannot be trusted. Hence, immunity to the main noise sources is crucial. 

\section{Robustness of the ground states to noise} 
To benefit from a long coherence time we need to be decoupled from the main noise sources. The most fidelity damaging noise source is the ambient magnetic field. Usually, to simulate spin-$1/2$ systems, the \Yb clock states are used. However, when simulating spin-1, the use of the Zeeman levels leaves the system vulnerable to magnetic field fluctuations. The second most important noise source is the fluctuations of the Rabi frequencies of the non-copropagating Raman beams that generate the spin-spin interaction $\Omega$, the carrier transition $\Omega_{car}$, and the driving fields that generate $\Omega'\sin\theta F_\alpha$. In order to counter these noise sources, we combine the continuous version of the dynamical decoupling technique and the unique quality of this model that makes it possible to conduct the experiment in a decoherence-free subspace. In our scheme, we use driving fields that refocus the noise in directions perpendicular to the final basis we operate with $(\overline{F_{z}})$. Specifically, these driving fields operate as dressing fields performing dynamical decoupling, thus, we are left with noise sources in the $z$ direction.  However, since the relevant ground states of the large $D$ and the Haldane phases belong to the decoherence free subspace, we are protected against these noise terms. 

The ground state of the large D phase is the topologically trivial state of a tensor product of $\left|\overline 0\right\rangle$, and the ground state of the Haldane phase \cite{Z2_Z2_1,Z2_Z2_2} has the same number of sites occupied by $\left|\overline 1\right\rangle$ and $\left|\overline {-1}\right\rangle$ , where  $\left|\overline 1\right\rangle$ $\left|\overline 0\right\rangle$ and  $\left|\overline {-1}\right\rangle$ are the eigenstates of $\overline{F_{z}}$ with eigenvalues $1,0,-1$ respectively. Therefore, these ground states are eigenstates of $\sum_j\overline{F^j_{z}}$, with a zero eigenvalue, such that our model is decoupled from this operation and thus from these noise sources. Thus, the quantum simulation operates in the decoherence free subspace.  For the same reason, all the neglected A.C. Stark shifts resulting in $\sum_j\overline{F^j_{z}}$ could have been dropped in the derivation.

\section{verification of the Haldane phase's ground state }
The ground states of the Haldane phase are characterized by: $(1)$ an excitation gap and exponentially decaying correlations of the local order $C_{f}^\alpha(i-j)=\left\langle S_\alpha^iS_\alpha^j \right\rangle$, where $\left\langle \right\rangle$ denotes the expectation value in the ground state, $(2)$ a nonvanishing nonlocal string order $O_{string}^{\alpha}\left(H\right)=\lim_{|i-j|\rightarrow   \infty}C_{st}^\alpha(i-j)$, where $C_{st}^\alpha(i-j)=\left\langle -S^{i}_{\alpha}\exp\left[i\pi\sum_{l=i+1}^{j-1}S^{l}_{\alpha}\right]S^{j}_{\alpha}\right\rangle$ is the string correlation function, $(3)$ a symmetry-protected double degenerate entanglement spectrum, obtained by dividing the systems into two parts, tracing out one of them and diagonalizing the reduced density matrix \cite{entanglement spectrum}.

Experimentally verifying the ground states of the Haldane phase can be accomplished by directly measuring the correlation functions of the local orders and the string order (signatures $(1)$ and $(2)$).
%Measuring experimentally the correlation functions of the local orders and the string order (signatures $(1)$ and $(2)$) can be done in the following way. 
For simplicity, suppose we want to measure $C_f^z(i-j)$ or equivalently $C_{st}^z(i-j)$. Experimentally we can only measure the dark singlet state $\ket{0}$ without the ability to distinguish between the bright triplet states $\ket{\pm 1}$, yet by single addressing we can rotate a chosen spin and measure the other states as well. Since $\left\langle S_z^iS_z^j \right\rangle= P^{i,j}_{1,1}+P^{i,j}_{-1,-1}-P^{i,j}_{1,-1}-P^{i,j}_{-1,1}$,%$\left\langle \ket{1}^i\bra{1}\ket{1}^j\bra{1}  \right\rangle+ \left\langle \ket{-1}^i\bra{-1} \ket{-1}^j\bra{-1} \right\rangle - \left\langle \ket{1}^i\bra{1}\ket{-1}^j\bra{-1}  \right\rangle - \left\langle \ket{-1}^i\bra{-1}\ket{1}^j\bra{1}  \right\rangle$
where $P^{i,j}_{a,b}$ is the probability to measure $\left\langle \ket{a}^i\bra{a}\ket{b}^j\bra{b}  \right\rangle$, 
four measurements would suffice \cite{actually}. The same holds for the string correlations, except here we have to count the bright triplet states of the intermediate spins, $\exp\left[i\pi\sum_{l=i+1}^{j-1}S^{l}_{z}\right]$. To measure the correlations in other basis, we first globally rotate the system to the desired basis, and then implement the same procedure.  

Full tomography is simply infeasible experimentally for increasingly long chains. However, in order to measure the entanglement spectrum (signature $(3)$), we can make tomography of a part of the system only. Namely, we can measure the reduced density matrix of this part containing a few spins, and diagonalizing it numerically. 

\section{A new approach for covering the whole positive $\lambda>0$ plane}
In the above scheme, the Ising-like $\lambda$ parameter is limited $0 \leq\lambda \leq 2$, and we can not cover the whole phase diagram, just like in the derivation of the previous paper \cite{Itsik}. This has to do with the $F_\beta^iF_\beta^j$ term, where $\alpha \perp \beta$, in the XX Hamiltonian (eq.\ref{XXHamiltonian}). In order to solve it, we have to suppress this limiting term; namely, we have to perform a MS like gate of spin-one, rather than the red side-band transition which results in the limiting XX Hamiltonian. Once we generate the $F_x^iF_x^j$ interaction term instead of the XX Hamiltonian, we can span the whole positive $\lambda$ plane of the phase diagram by using the same $\lambda-$generating trick as above.  

As was suggested by Senko {\it et al} \cite{Senko_Haldane}, the straight-forward way to generate the MS transitions is by applying additional beat frequencies to drive the blue sideband transitions in addition to the red sideband ones, based on Ref. \cite{SM}. Yet, in this section, we would like to show another way to realize the MS Hamiltonian, based on the Bermudez et al. gate proposal \cite{Bermudez gate}. Here, we use the carrier transitions that were used to generate $\Omega'\sin\theta F_\alpha$ in the above derivation, in order to suppress the limiting term, and generate the $F_x^iF_x^j$ interaction (Ising Hamiltonian).
%apply additional driving field for driving the carrier transitions  $\left\vert -1\right\rangle \rightarrow \left\vert 0 \right\rangle $ and $ \left\vert 1 \right\rangle \rightarrow \left\vert 0 \right\rangle $. It can be achieved using co-propagating Raman beams, or using resonant microwave driving fields. 
Thus, in the rotating frame of the bare state energy levels, the carrier transitions yield
\begin{equation}
H_{car}=\frac{\Omega_{car}}{\sqrt{2}}\sum_j F_\alpha^j,
\label{micro}
\end{equation} 
where $\alpha$ is in the XY plane, and is determined by the relative phase of the two carrier transitions (fig. \ref{F_a1} b). For simplicity, we assume that $\alpha=x$ corresponding to a vanishing initial phase between the carrier transitions. These transitions dress our qutrits, and have significant implications in terms of suppressing the magnetic field noise. Moving to the dressed state basis can be thought of as a $-\pi/2$ rotation about the $y$ axis; thus, $F_x, F_y, F_z $ in the bare state basis are transformed to $ S_z,S_y,-S_x$ in the dressed state basis, respectively. In the rotating frame of the dressed state energy structure (eq. \ref{micro}), the red sideband transition (eq. \ref{red}) becomes:
 \begin{equation}
\begin{split}
 \label{red2} 
 H_{red}=  \sum_{n,j}\frac{i \Omega\eta_{j,n}}{2\sqrt{2}} \left(  \left[ S_z^j+ \frac{S_+^j}{2} e^{i\frac{\Omega_{car}}{\sqrt{2}}t}- \frac{S_-^j}{2} e^{i\frac{\Omega_{car}}{\sqrt{2}}t} \right] e^{i\delta t} -h.c \right)\\
 \left( b_{n}^\dagger e^{i\nu_n t} +h.c \right) .
\end{split}
\end{equation}
If $ \delta - \nu_n \ll \Omega_{car}/\sqrt2 $ we can neglect the fast rotating terms; thus we are left with the MS Hamiltonian for qutrits:
 \begin{equation}
\begin{split}
 \label{red3} 
 H_{red}=  \sum_{n,j}\frac{i \Omega\eta_{j,n}}{2\sqrt{2}} \left(   S_z^j e^{i\delta t} -h.c \right)
 \left( b_{n}^\dagger e^{i\nu_n t} +h.c \right) ,
\end{split}
\end{equation}
resulting in the Ising Hamiltonian for spin-one systems, in the second order perturbation theory, if $  \Omega\eta_{j,n}/{2\sqrt{2}} \ll \delta - \nu_n$:
 \begin{equation}
     \label{IsingHamiltonian}
 H_{Ising}=\sum_{i\le j}J_{ij}^{eff}\left(S_{z}^{i}S_{z}^{j}\left(1-\delta_{i,j}\right)+\frac{
 {\left(S_{z}^{j}\right)^{2}}}{2}\delta_{i,j}\right),\\
\end{equation}
This looks like we are not  proceeding with the derivation of eq. \ref{HaldaneHamiltonian}, but rather are going backwards. We have already had the XX Hamiltonian (eq. \ref{XXHamiltonian}), and with additional effort (eq. \ref{micro}), we only obtain the Ising Hamiltonian.  However, using the trick for generating the $\lambda-$like term, we also generate the XXZ Hamiltonian, this time covering the whole positive $\lambda$ phase diagram, unlike the previous derivation. In the following, we pursue a tunable $D$ term, and a tunable $\lambda$ term. 
%%%%%%%%%%%%%%%%%
% talk about phase match between the lasers and the microwaves. 
%talk about the initial phase determining the dressed state basis.
%%%%%%%%%%%%%%%%%

{\it Generation of the control anisotropy $D$-term ---} As before, generating the tunable $D$-term can be done in two ways: generating an A.C. Stark shift, or imposing detunings. To generate the detuned transitions to implement the first method we can apply an additional $\Delta_D$ detuned microwave driving field, corresponding to the transition between the two zero states of the triplet and the singlet, $\left|0'\right\rangle\longleftrightarrow\left|0\right\rangle$. In a similar way, we can obtain the same result by applying co-propagating Raman beams corresponding to the same transition (fig.\ref{D_fig_1} a).  We then follow the previous steps of derivation: by first transforming to the dressed state basis, where $\left\vert 0 \right\rangle \rightarrow \left( \left\vert u \right\rangle - \left\vert d \right\rangle \right)/\sqrt2$, and then moving to the rotating frame of $\frac{\Omega_{car}}{\sqrt{2}}S_{z}$, we obtain
\begin{equation}
\frac{\Omega_D}{2\sqrt2} \left( \left\vert u \right\rangle \left\langle 0' \right\vert e^{i (\Delta_D+\frac{\Omega_{car}}{\sqrt2} ) t } -\left\vert d\right\rangle \left\langle 0' \right\vert e^{i(\Delta_D-\frac{\Omega_{car}}{\sqrt2}) t } \right) + h.c.
\label{A.C2}
\end{equation} 
In the second perturbation theory, assuming $ \frac{\Omega_D}{2\sqrt2},\frac{\Omega_{car}}{\sqrt2} \ll \Delta_D $, and $  \frac{\Omega_D^2}{8\Delta_D} \ll \sqrt2\Omega_{car} $, only the A.C. Stark shifts survive, whereas the other off-diagonal terms coming from the Raman transitions between $\left\vert u\right\rangle\longleftrightarrow\left\vert d \right\rangle$ are suppressed in the RWA. Thus, we are left with the desired $D'(S_z)^2$ term, setting $D'=\frac{\Omega_D^2}{8\Delta_D}.$

Generating a tunable $D$-term can alternatively be achieved by regarding the $\left|0\right\rangle$ bare state energy level as $2D'$-shifted corresponding to its real value in the hyperfine structure (fig.\ref{D_fig_1} b). That is to say, we impose $2D'$-detuning on all the driving fields generating the Ising Hamiltonian (eq. \ref{IsingHamiltonian}). Therefore, we are left with the term $2D' \left\vert 0\right\rangle \left\langle 0 \right\vert$. Transforming to the dressed-state basis and moving to the interaction picture with respect to $\frac{\Omega_{car}}{\sqrt{2}}S_{z}$ as was mentioned above, all the off-diagonal terms are suppressed using the RWA, if we assume that $D'\ll\sqrt{2}\Omega_{car}$. Once again, we are left with  $D'(S_z)^2$. %In order to tune this term we have to continuously shift all the driving fields transitions \cite{xx1}. Note that the anisotropy D-term is slightly different from these D' terms, since the model is $\theta$ rotated while generating $\lambda$-term. 

%\begin{figure} 
%\centering 
% \includegraphics[width=0.5\textwidth]{2}
 % \caption{{\bf Generating the D-term} Generating the D-term in Eq. \ref{HaldaneHamiltonian} is done in two ways: 1) (Right side) by applying two additional co-propagating Raman beams, corresponding to a $\Delta_D$ detuned transition between $\left|0'\right\rangle\longleftrightarrow\left|0\right\rangle$, we are left with $\frac{\Omega_D^2}{4\Delta_D}\left\vert 0\right\rangle \left\langle 0 \right\vert $, due to the A.C. Stark shift, in the second perturbation theory. 2) (Left side) by regarding the $\left\vert 0 \right\rangle$ state as a 2D'-shifted, and moving to the i.p.w.r.t the bare-state 2D'-shifted energy structure, we are left with $2D'\left\vert 0 \right\rangle \left\langle 0 \right\vert$. In both ways we obtain $D'S_z^2 $ after transforming to the dressed-state basis.}
% \label{D}
%\end{figure}
\begin{figure} 
\centering 
 \includegraphics[width=0.5\textwidth]{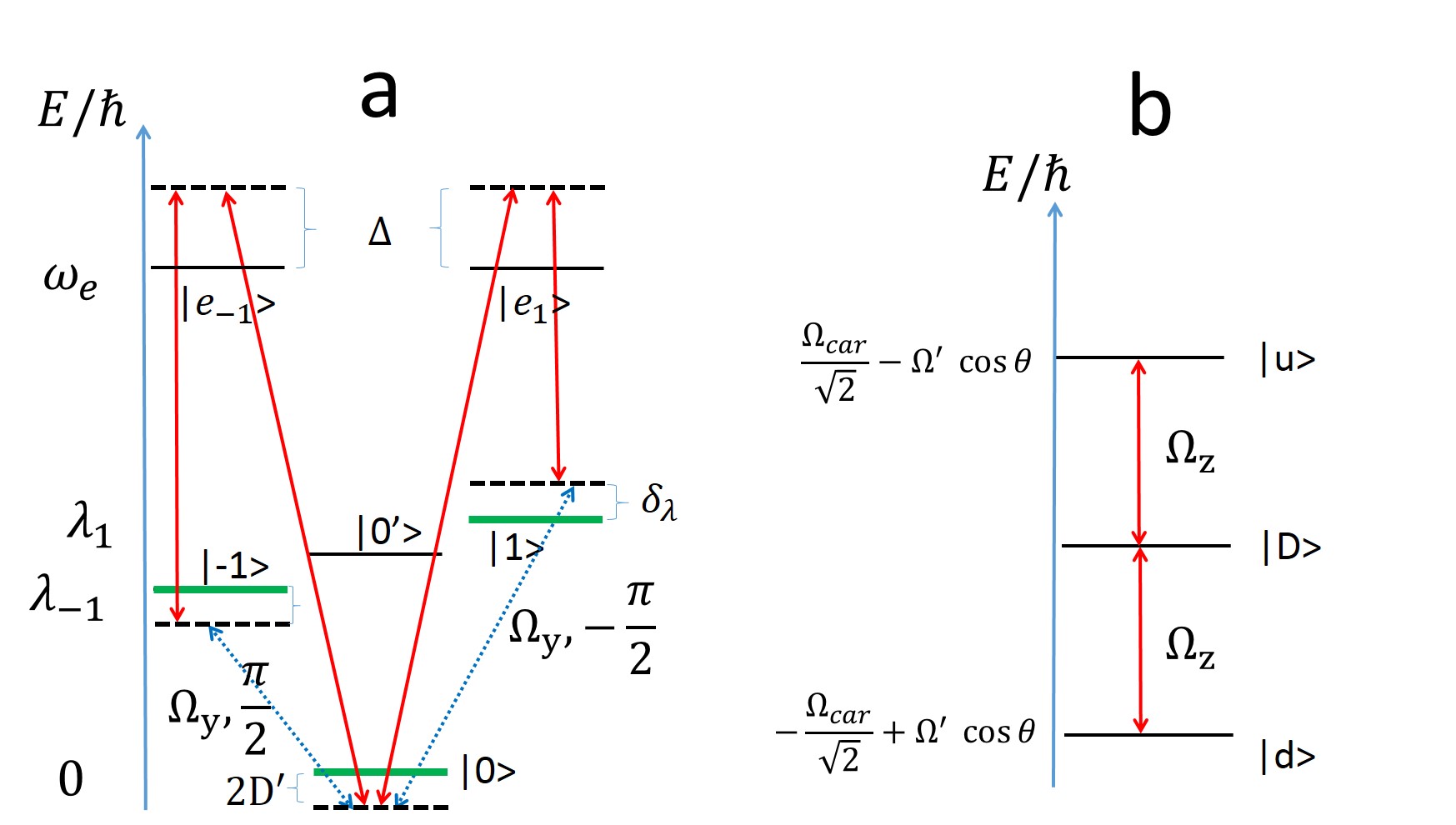}
  \caption{{\bf Generating $\Omega'S_\alpha\sin\theta$ for the $\lambda$-generating trick.} There are two ways to realize this term of the $\lambda$-generating trick: (1) (a.) to engineer $\Omega' S_y \sin\theta $, we apply additional co-propagating laser Raman beams, which perform the two Raman transitions between $\left|\mp1\right\rangle \leftrightarrow\left|0\right\rangle$ with $\pm\delta_\lambda=\pm\left(\frac{\Omega_{car}}{\sqrt{2}}-\Omega'\cos\theta\right)$ detunings, $\pm\frac{\pi}{2}$ initial phases and the same effective Rabi frequency $\Omega_y=\sqrt{2}\Omega'\sin\theta$ respectively. (2) (b.)  to engineer $-\Omega' \sin\theta S_x  $ we apply a z-polarized radio-wave driving field, on resonance with the dressed-state energy structure (fig.\ref{F_a1} b), and with a Rabi frequency of $\Omega_z=2\Omega'\sin\theta$. %$\cdot F_z \cos{\frac{\Omega}{\sqrt2} t}$. This is transformed to $ -2\Omega'\sin\theta S_x \cos{\frac{\Omega}{\sqrt2} t} $ in the dressed state basis. 
Regardless of method, we obtain the desired term by moving to the rotating frame of the dressed energy structure, and using RWA while assuming $\Omega' \ll \sqrt2\Omega$.}
 \label{L}
\end{figure}
 
  {\it Generation of the Ising-like $\lambda$-term ---} To implement the $\lambda$-generating trick; i.e., adding a spin operator term $\theta$-rotated from the dressed $z$ axis, which is $\Omega' S_{z,\theta}=\Omega' \left( S_z \cos\theta +S_{\alpha}\sin\theta \right)$, we generate each component separately. The first $\Omega'S_{z}\cos\theta$ term is easily produced since it can be set aside from the carrier transition (eq. \ref{micro}), such that instead of eq. \ref{micro}, $\left(\frac{\Omega_{car}}{\sqrt{2}}-\Omega'\cos\theta \right)S_{z}$ is used as the dressing term. Regarding the second term $\Omega' S_{\alpha}\sin\theta$ there are two alternatives. Choosing $\alpha=y$, the term $\Omega'S_{y}\sin\theta$ is produced by using additional microwave driving fields, corresponding to the two transitions between $\left|\mp1\right\rangle \leftrightarrow\left|0\right\rangle$ with $\pm\left(\frac{\Omega_{car}}{\sqrt{2}}-\Omega'\cos\theta\right)$ detunings, $\pm\frac{\pi}{2}$ initial phases and the same Rabi frequency $\Omega_y=\sqrt{2}\Omega'\sin\theta$ respectively (fig.\ref{L} a). Similarly, we can obtain the same result with co-propagating laser Raman beams corresponding to these transitions.

 Choosing $\alpha=x$, the $-S_{x}\sin\theta$ term, is simply obtained by applying an additional z-polarized radio-frequency (RF) driving field, $\Omega_{z} F_z \cos \left(\left(\frac{\Omega_{car}}{\sqrt{2}}-\Omega'\cos\theta\right) t\right)$, with $\Omega_{z}=2\Omega'\sin\theta$ (fig.\ref{L} b). Thus, by moving to the rotating frame of the bare-state energy structure, the RF driving field is not affected. Finally, for both $\alpha=x,y$, the desired terms are obtained in the rotating frame corresponding to $\left(\frac{\Omega_{car}}{\sqrt{2}}-\Omega'\cos\theta\right)S_z$, using the RWA where we assume that $\Omega' \ll \sqrt2\Omega_{car}$. %For simplicity, these additional transitions for generating $\Omega'\cos\theta S_\alpha$ for both $\alpha=x,y$  will be named RF transitions.
 
By applying a $\theta$ rotation around the perpendicular axis, such that $S_{z,\theta}$ is transformed into $\overline{S_{z}}$, the new spin operators and the double dressed state basis are $\theta$ rotated as well.   %the operators are transformed to: $S_{z,\theta}\rightarrow S_{z}$, $S_{z}\rightarrow S_{z}\cos\theta+S_{x}\sin\theta$, $S_{x}\rightarrow S_{x}\cos\theta-S_{z}\sin\theta$, $S_{y}\rightarrow S_{y}$, and the new dressed-state basis is $\theta$ rotated as well. 
  If we move to the interaction picture with respect to  the new term we have built $\Omega'\overline{S_{z}}$ and use the RWA where we require $\frac{(\Omega\eta_{j,n})^2}{8(\delta-\nu_n)}\ll\Omega' \ll \delta-\nu_n $, we end up with the following effective Hamiltonian: 
  \begin{equation}
  \begin{split}
\label{Heff_final2}
 H_{eff}=\sum_{i < j} J_{ij}^{eff}\left\lbrace\left( \overline{S_{x}^{i}}\overline{S_{x}^{j}}+\overline{S_{y}^{i}}\overline{S_{y}^{j}} \right)\frac{\cos^2\theta}{2} + \overline{S_{z}^{i}}\overline{S_{z}^{j}}\sin^2\theta\right\rbrace\\
+\sum_i\left(D'-\frac{J_{ii}^{eff}}{2}\right)\left(\frac{\cos^2\theta}{2} -\sin^2\theta \right)(\overline{S_z^i})^2.
 \end{split}
 \end{equation}
Therefore, we have obtained the required spin-one  {\it XXZ} AFM Hamiltonian, while covering the whole positive $\lambda \geq 0$ plane in the phase diagram. Experimentally, the following parameters are realistic: $\nu_n=2\pi \cdot 5$MHz, $\delta=2\pi  \cdot 5.1$MHz, $\Omega_{car}=2\pi  \cdot 1$MHz,  $\Omega'=2\pi  \cdot  10$kHz, $\Omega=2\pi  \cdot  500$kHz ,$\eta_{n,j}\approx 0.14$,which gives $J_{ii}^{eff}=2\pi \cdot1$kHz  , such that $0<D<2\pi 10$kHz.

\section{experimental realization of moving to the interaction picture}
In this scheme of generating the {\it XXZ-D} Hamiltonian, we move to three interaction pictures: the first one is with respect to the bare energy gap of the qutrit, which yields the same first interaction unitary $U_0(\tau)$ (eq. \ref{U1}) of the first approach.
%\begin{equation}
%U_0(\tau)=\exp\left(-i\left[(\omega_0-D') \left\vert 0 \right\rangle \left\langle 0 \right\vert  + \lambda_1\left\vert 1 \right\rangle \left\langle 1 \right\vert -\lambda_1\left\vert -1 \right\rangle \left\langle -1 \right\vert  \right]\tau \right)  
%\end{equation} 
%assuming we use the detuning approach for generating the D term. 
The second interaction picture is with respect to the microwave dressed state energy gap, giving rise to the following interaction unitary:
\begin{equation}
U_1(\tau)=\exp\left(-i \left[\frac{\Omega_{car}}{\sqrt{2}}-\Omega'\cos\theta\right] F^j_x\tau\right)
\end{equation}
and the last interaction picture is with respect to a superposition of the $\theta-$rotated term used for generating the $\lambda-$ generating trick. It results in the following interaction unitary:
\begin{equation}
U_2(\tau)=\exp\left(-i \Omega'\left[\cos\theta F_x+\sin\theta F_\alpha \right] \tau\right),
\end{equation}
with $\alpha=z,y$.
Moving to these interaction pictures results in the effective Hamiltonian (eq. \ref{Heff_final2}), which yields the simulated evolution $U_{I_2}(\tau)$. 

As was discussed above, in order to move to the interaction frame, we first have to shut down the effective Hamiltonian. This is done by blocking the two non-copropagating Raman beams (setting $\Omega=0$), and eliminating the $D'$ term for $t_2$ time duration, such that $U_2(\tau+t_2)=\mathbb{I}$.
%Recall that we have suggested two ways to generate it, either using a $\Delta_D-$detuned transition that results in an A.C Stark shift resulting in the $D'$ term (eq. \ref{A.C}), or by imposing a $2D'-$detuning to all the relevant driving fields (the non-copropagating Raman transitions and the carrier transitions). Therefore, we can block the driving fields that generate this term with an A.C Stark shift, or by canceling the detuning of the relevant driving fields, if the second approach is taken. Shutting down the effective Hamiltonian should take $t_2$ such that $U_2(\tau+t_2)=\mathbb{I}$.

The next stage is countering $U_1$. For that purpose, we have to shut down the driving fields responsible for $U_2$. Specifically we have to block the driving fields that generate $\Omega'\sin\theta F_\alpha$, and to change the Rabi frequency of the carrier transitions to $\Omega_{car}/\sqrt{2}-\Omega'\cos\theta$. This stage should take $t_1$, such that $U_1(\tau+t_2+t_1)=\mathbb{I}$. As before, there is no need to counter $U_0$, and we can measure in any basis we desire.

%There is no need to counter $U_0$ since it yields only an unmeasured phase. Leaving the experiment in this point results in measuring the state in the $F_z$ basis. Now, if we would like to measure in the simulated basis, namely $\overline{S_{z}}= \cos\theta F_x+\sin\theta F_\alpha $ for $\alpha=z,y$, we have to rotate the system around a perpendicular axis. Suppose that using the RF driving field $\alpha=z$, then we have to rotate the system with $F_y$ operation. This can be achieved after countering the both $U_2$ and $U_1$, by a $\pi/2$ phase change of the microwave driving field, such that instead of operating with $F_x$ the microwave driving field will operate with $F_y$. In a similar way we can measure in any basis we choose.

%In the second approach of moving experimentally to the interaction picture, we can calculate the total interaction unitary $U_0(\tau)U_1(\tau)U_2(\tau)$ by quantum engineering its hermitian conjugate separately.

Similarly to the above scheme, during the adiabatic quantum simulation experiment, we counter the main noise sources - the ambient magnetic field fluctuations, and the Rabi frequency fluctuations - using a combination of the continuous dynamical decoupling technique with decoherence-free subspace.

\section{Summary} 
We have discussed how to quantum engineer the spin-one XXZ-D AFM Hamiltonian with two schemes. The first covers $0\leq\lambda \leq 2$, whereas the second covers the whole positive $\lambda \geq 0$ plane in the phase diagram. It enables us to explore the regime of the Heisenberg Hamiltonian of integer spin systems where the Haldane phase resides. We have explained how to adiabatically generate this non-trivial topological phase, starting from the large $D$ phase with its trivial ground state. During the adiabatic path, the ground states are robust to the fluctuations in the magnetic field and Rabi frequencies, 
% main noise sources, the magnetic noise, and the fluctuations in the Rabi frequencies, 
and belong  to a decoherence free subspace. This permits a longer adiabatic path with higher fidelities. We have also shown how the Haldane phase can be verified with simple experimental measurements. This proposal may thus constitute an important step towards exploring topological phases with trapped ions.%, paving the way towards realizing highly frustrated spin liquid phases in higher dimensional ion lattices.

%This proposal may thus constitute an important step towards exploring topological phases in highly frustrated models in higher dimensions with trapped ions.

{\em Acknowledgements ---} We thank Erez Berg, Dror Orgad, Aaron Lee, Jacob Smith, Crystal Senko, and Alexey Gorshkov for useful discussions and acknowledge the support of the European Commission (STREP EQuaM Grant Agreement No. 323714). Z.-X.G. acknowledges the support of AFOSR, NSF PIF, the ARO, NSF PFC at the JQI, the ARL, and the AFOSR MURI. A. R. acknowledges the support of the Israel Science Foundation(grant no. 1500/13).
  \\


\begin{references}
\bibitem{actually}
We can measure the order correlations with three measurements: $P^{i,j}_{1,1}, P^{i,j}_{-1,-1}$ and $P^{i,j}_0$, where the latter is the probability to measure at least one of the ions ($i$ or $j$) in the $\ket{0}$ state. This uses the fact that $P^{i,j}_{1,-1}+ P^{i,j}_{-1,1} = 1- P^{i,j}_{1,1}- P^{i,j}_{-1,-1}-P^{i,j}_0$.
\end{references}
\end{document}